\documentclass[aps,preprint,preprintnumbers,amsmath,amssymb,nofootinbib]{revtex4}
\usepackage{amssymb}
\usepackage{lscape}
\usepackage{amsmath}
\usepackage{bm}
\usepackage{graphicx}
\usepackage{epsfig}
\begin{document}

\title{Large-scales solitonic back-reaction behavior in power-law inflation and its relationship with dark energy}
\author{$^{2}$ Juan Ignacio Musmarra\footnote{jmusmarra@mdp.edu.ar}, $^{1,2}$ Mariano Anabitarte\footnote{anabitar@mdp.edu.ar},  $^{1,2}$ Mauricio Bellini
\footnote{{\bf Corresponding author}: mbellini@mdp.edu.ar} }
\address{$^1$ Departamento de F\'isica, Facultad de Ciencias Exactas y
Naturales, Universidad Nacional de Mar del Plata, Funes 3350, C.P.
7600, Mar del Plata, Argentina.\\
$^2$ Instituto de Investigaciones F\'{\i}sicas de Mar del Plata (IFIMAR), \\
Consejo Nacional de Investigaciones Cient\'ificas y T\'ecnicas
(CONICET), Mar del Plata, Argentina.}

\begin{abstract}
We study the behavior of large-scale (cosmological) modes of back-reaction effects during inflation. We find that the group of modes which describes the very large-scale fluctuations of energy density during inflation due to back-reaction effects evolve in phase between them, but there is a tear of these modes with respect to the other modes that describe astrophysical scales. This effect could be the origin for the large-scale homogeneity and isotropy of the universe and could be a manifestation of the existence of dark energy, which is responsible for the accelerated expansion of the universe.
\end{abstract}
\maketitle

\section{Introduction and motivation}

In cosmology, the inflationary theory of the universe provides a well established physical mechanism that can explains why the universe is so homogeneous, isotropic an flat. It also provides a mechanism to generate the primordial energy density fluctuations on cosmological scale, that help to explain the formation of structure of the universe\cite{infl1,infl2,infl3}. The primordial scalar metric perturbations, related to the energy density fluctuations, are the seeds that drive the formation of large scale structure, which gave origin to today's galaxies. These fluctuations that today are larger than a thousand times the size of a typical galaxy, during inflation where much larger than the size of the casual horizon, and have been being tested by past and current observations of cosmic microwave background (CMB)\cite{cmb}.

According to the inflationary theory, a scalar field minimally coupled to gravity, called inflaton, under the effects of a quasi constant potential, causes an accelerated expansion of the early universe. During this period we can consider the potential energy density to be dominant, therefore the kinetic energy can be neglected. This is known as the slow roll condition for the inflaton field dynamics\cite{inflaton}. In this case we study back-reaction effects within this model. This arise because we are studying a non-perturbative method for the scalar metric fluctuations. This quantum vacuum fluctuations are continuously generated in sub-Hubble scales, and as the wavelength of these fluctuations exit the Hubble radius the modes of the fluctuation get squeezed and become the seeds for the observed inhomogeneities in the anisotropy and matter distribution. At that time, the quantum fluctuations of the scalar field and of the metric undergo a semi-classical transition.

In this work we will study the dynamics of the gauge-invariant scalar metric fluctuations using a non-perturbative method, the Relativistic Quantum Geometry (RQG), wich describes the dynamics of the geometrical back-reaction effects as a departure of a background Riemannian spacetime with the help of a quantum geometrical scalar field. The dynamics of the geometrical scalar field $\sigma $ is defined on a Weyl-integrable manifold that preserves the gauge-invariance under the transformations of the Einstein's equations\cite{rb1,rb2}, that involves the cosmological constant. In particular, we are interested in providing a description of the possible solitonic behavior of $\sigma$ on very large-scales and the link with dark energy. A soliton is a self-reinforcing solitary wave packet that maintains its shape while it propagates. A description of Ricci solitons in Einstein-Scalar field theory and fluid spacetimes was recently studied in \cite{she,she1}. They are caused by a cancellation of nonlinear and dispersive effects in the medium so that the soliton propagates with coherence in their modes. For this reason its existence could be interesting to explain the very large-scale coherence of back-reaction effects and its relation with dark-energy during inflation\cite{wet}, which is a topic that remains unstudied.

\section{inflation and back-reaction}

In this section we shall study the dynamics of a power-law inflationary model with back-reaction effects included. These effects take into account the metric reaction to the fluctuations of the inflaton field, and are considered as geometrical. It is very important for these effects to be described in a nonperturbative manner, in order for them to be able to describe strong field departures of the background space-time.

\subsection{Background power-law inflationary dynamics}

In order to study the dynamics of back-reaction effects during inflation, we consider the background metric given by a Friedman-Robertson-Walker (FRW) metric
which describes a spatially flat, homogenous and isotropic universe on cosmological scales
\begin{equation}\label{back}
d\hat{S}^2 = \hat{g}_{\mu\nu} d\hat{x}^{\mu} d\hat{x}^{\nu}= d\hat{t}^2 - a^2(\hat{t}) \hat{\eta}_{ij} d\hat{x}^i d\hat{x}^j,
\end{equation}
where the {\em hat} denotes that the metric tensor is defined over a semi-Riemannian manifold. We shall consider a power-law expansion of the universe during inflation. This means that the power of the expansion must be large enough: $p\gg 1$, and the scale factor of the universe must be defined as
\begin{equation}\label{sca}
a(\hat{t})=a_0\left(\frac{\hat{t}}{t_0}\right)^p,
\end{equation}
for a Hubble parameter $H\left(\hat{t}\right)=p/\hat{t}$. Here, $t_0$ is the initial value of the cosmic time. The background dynamics of the scalar field that generates the inflation of the universe, called the inflaton field, is
\begin{equation}
\ddot\phi + 3 H(\hat{t})\, \dot\phi + V'(\phi) =0,
\end{equation}
such that the scalar potential $V(\phi)$ written in terms of the inflaton field, is
\begin{equation}
V(\phi) = \frac{3  H^2_0}{\kappa} \left(1 - \frac{2}{3 \kappa \phi^2_0}\right) e^{2 (\phi/\phi_0)},
\end{equation}
where $\kappa=8\pi G$, $p = (\kappa/2) \phi^2_0$, $a_0$ is the initial value of the scale factor, $\phi_0$ the value of the inflaton field when inflation starts, and $H_0$ is a constant with the same units of the Hubble parameter\cite{prd96}.

\subsection{Back-reaction effects and dark energy}

The study of nonperturbative back-reaction effects was introduce in\cite{rb1}, with a displacement from a Riemann manifold, to a manifold described with connections\footnote{We denote $\sigma_{\alpha} \equiv \sigma_{,\alpha}$.}
\begin{equation}\label{gama}
\Gamma^{\alpha}_{\beta\gamma} = \left\{ \begin{array}{cc}  \alpha \, \\ \beta \, \gamma  \end{array} \right\}+\,\sigma^{\alpha} \hat{g}_{\beta\gamma}.
\end{equation}
Here $\delta{\Gamma^{\alpha}_{\beta\gamma}}=\,\sigma^{\alpha} \hat{g}_{\beta\gamma} $ describes the displacement of the extended manifold with respect to the Riemannian background. The Riemann manifold is described by the Levi-Civita symbols in (\ref{gama}). As can be demonstrated, the covariant derivative of the metric tensor on the extended manifold is nonzero: $ g_{\alpha\beta|\gamma}= -\left[\sigma_{\alpha}\,g_{\gamma\beta}+\sigma_{\beta}\,g_{\gamma\alpha}\right]$, but of course, it is null on the Riemann manifold: $\nabla_{\gamma} g_{\alpha\beta}=0$. From the action's point of view, the scalar field $\sigma(x^{\alpha})$ is a generic geometrical transformation that leaves invariant the action\cite{rb2}:
\begin{equation}\label{aac}
{\cal I} = \int d^4 \hat{x}\, \sqrt{-\hat{g}}\, \left[\frac{\hat{R}}{2\kappa} + \hat{{\cal L}}\right] = \int d^4 \hat{x}\, \left[\sqrt{-\hat{g}} e^{-2\sigma}\right]\,
\left\{\left[\frac{\hat{R}}{2\kappa} + \hat{{\cal L}}\right]\,e^{2\sigma}\right\},
\end{equation}
so that the following must be complied\footnote{Here, $\delta\sigma = \sigma_{\mu} dx^{\mu}$ is an exact differential and $\hat{v}=\sqrt{-\hat{ g}}$ is the volume of the Riemannian manifold. Of course, all the variations are in the Weylian geometrical representation, and assure us gauge invariance because $\delta {\cal I} =0$.}
\begin{equation}\label{var}
-\frac{\delta v}{\hat{v}} = \frac{\delta \left[\frac{\hat{R}}{2\kappa} + \hat{{\cal L}}\right]}{\left[\frac{\hat{R}}{2\kappa} + \hat{{\cal L}}\right]}
= 2 \,\delta\sigma.
\end{equation}
A metric that takes into account back-reaction effects on the background space-time (\ref{back}), is\cite{ab}
\begin{equation}\label{mm}
d{S}^2 =   e^{2\sigma}\,d\hat{t}^2 - a^2(\hat{t}) {\delta}_{ij}\,e^{-2\sigma}\, d\hat{x}^i d\hat{x}^j,
\end{equation}
which is related to a volume of the manifold given by the squared root of its determinant $v=\sqrt{-\hat{ g} }\,e^{-2\sigma}$. In particular, this choice is very interesting because, as was demonstrated in\cite{lid}, leaves invariant the equation of state of the system. If $(\hat{P}, \hat{\rho})$ are the background pressure and energy density and $({P}, {\rho})$ are the same quantities on the extended manifold, it can be demonstrated that $\delta f=-\sigma + \frac{1}{3(\omega+1)}\,{\rm ln}[\rho/\hat{\rho}]$ is a gauge-invariant quantity representing the non-linear extension of the curvature perturbation for adiabatic fluids on uniform energy density hypersurfaces at superhorizon scales\cite{lid,ab}, with $\omega=P/\rho=\hat{P}/\hat{\rho}$. From the point of view of the Einstein-Hilbert action, we are considering that the flux along a closed hypersurface of the background (Riemannian) spacetime is nonzero, so that
\begin{equation}
g^{\alpha\beta} \frac{\delta R_{\alpha\beta}}{\delta S} = - \frac{\Lambda}{2} \sigma_{\gamma} \,\hat{U}^{\gamma},
\end{equation}
where $\hat{U}^{\gamma}\left|B\right>= \frac{\delta x^{\gamma}}{\delta S}\left|B\right>$ are the relativistic components of the $4$-velocities on the extended manifold defined by the connections (\ref{gama}), evaluated on the quantum state: $\left|B\right>$\cite{rb1,rb2}, and $\Lambda(\sigma;\sigma_{\alpha})$ can be considered as a functional such that it is possible to define a quantum action: ${\cal W} = \int d^4 x \, \sqrt{-g} \,\, \Lambda(\sigma, \sigma_{\alpha})$. In that case, the dynamics of the geometrical field given by the Euler-Lagrange equations, after imposing $\delta W=0$, is\footnote{The reader can see a more detailed explanation in \cite{rb1}}
\begin{equation}\label{EL}
\frac{\delta \Lambda}{\delta \sigma} - \hat{\nabla}_{\alpha} \left( \frac{\delta \Lambda}{\delta \sigma_{\alpha}}\right) =0.
\end{equation}
The equation (\ref{EL}) provides the dynamics of the geometrical scalar field $\sigma$. This geometrical field is the responsible for the Einstein tensor perturbation related to
$\Lambda$ through the expression: $\delta G_{\alpha\beta}=-g_{\alpha\beta} \,\Lambda$, where
\begin{equation}
\delta G_{\alpha\beta} = \nabla_{\beta} \sigma_{\alpha} +\sigma_{\alpha} \sigma_{\beta} + \frac{1}{2} \,g_{\alpha\beta}
\left[ \Box\,\sigma + \sigma_{\mu} \sigma^{\mu} \right],
\end{equation}
where $\Box\,\sigma=\nabla_{\alpha} \sigma^{\alpha}=0$, and\cite{rb1}
\begin{equation}
\nabla_{\beta} \sigma_{\alpha}=\sigma_{\alpha,\beta} -\left\{ \begin{array}{cc}  \epsilon \, \\ \beta \, \alpha   \end{array} \right\} \,\sigma_{\epsilon},
\end{equation}
with the quantization
\begin{equation}\label{con}
\left[\sigma(x),\sigma^{\alpha}(y) \right] =- i \Theta^{\alpha}\, \delta^{(4)} (x-y), \qquad \left[\sigma(x),\sigma_{\alpha}(y) \right] =
i \Theta_{\alpha}\, \delta^{(4)} (x-y),
\end{equation}
such that $\Theta^{\alpha} = \hbar\, \hat{U}^{\alpha}$. In a cosmological gauge, where the background metric is described by a FRW one, the $4$-velocities $\hat{U}^{\alpha}$ take values $(1,0,0,0)$, because the relativistic observer is moving in a co-moving frame.

\subsection{Energy density fluctuations due to back-reaction effects}

The geometrical scalar field $\sigma$ can be expressed as a Fourier expansion
\begin{equation}\label{four}
\sigma(\vec{\hat{x}},\hat{t}) = \frac{1}{(2\pi)^{3/2}} \int \, d^3k \, \left[ A_k \, e^{i \vec{k}.\vec{\hat{x}}} \xi_k(\hat{\hat{t}}) + A^{\dagger}_k \, e^{-i \vec{k}.\vec{\hat{x}}} \xi^*_k(\hat{\hat{t}}) \right],
\end{equation}
where $A^{\dagger}_k$ and $A_k$ are the creation and annihilation operators
\begin{equation}
\left<B\left|\left[A_k, A^{\dagger}_{k'}\right]\right|B\right> = i\, \delta^{(3)}\left(\vec{k}-\vec{k'}\right),
\end{equation}
where $\left|B\right>$ is the quantum state defined on the Riemannian background (curved) space-time, in the Heisenberg representation. In power-law inflation the background scale factor $a(\hat{t})$ is given by (\ref{sca}). The dynamics for $\sigma$ is governed by the equation
\begin{equation}
\ddot\sigma + 3 \frac{\dot{a}}{a} \dot\sigma - \frac{1}{a^2} \nabla^2 \sigma =0.
\end{equation}
The back-reaction energy density fluctuations during inflation were calculated in a previous work, using RQG\cite{rb1}
\begin{equation}
\left<B\left|\frac{1}{\hat{\rho}} \frac{\delta \hat{\rho}}{\delta S}\right|B\right> = - 2 \, \dot{{\sigma}},
\end{equation}
where the {\em dot} denotes the derivative with respect to $t$. Here, $\dot{\sigma} = \left< (\dot{\sigma})^2 \right>^{1/2}$, such that
\begin{equation}
\left< \dot{\sigma}^2 \right> = \frac{1}{(2\pi)^{3}} \, \int\, d^3k  \left|\dot{\xi}_k \right|^2.
\end{equation}
The modes of the field $\sigma$: $\xi_k$, must be restricted by the normalization condition:
$({\dot\xi}_k^*) \xi_k - (\dot{\xi}_k) \xi^*_k= \frac{i}{a^3}$, in order for the field $\sigma$ to be quantized by using the expressions (\ref{con}), and the solution results to be
\begin{equation}
\xi_k(\hat{t}) = \sqrt{\frac{\pi}{4 (p-1)}} \hat{t}^{-(3p-1)/2} \,{\cal H}^{(2)}_{\nu}[y(\hat{t})],
\end{equation}
with $y(\hat{t})= { k \,\hat{t}^{(1-p)} \over \beta (p-1)}$ and $\beta ={a_0\over \hat{t}^p_0}$. The time derivative of the temporal modes in power-law inflation is
\begin{equation}
\dot{\xi}_k(\hat{t}) = \frac{1}{2} \sqrt{\frac{\pi}{p-1}} \hat{t}^{-2p} \left[ (1-3p) {\cal H}^{(2)}_{\nu}\left[y(\hat{t})\right] + \frac{k}{\beta} {\cal H}_{\nu_1}\left[y(\hat{t})\right]\right],
\end{equation}
where $\nu_1= \frac{5p-3}{2(p-1)}$. On large (cosmological) scales, the argument of the Hankel functions is very small: $y(\hat{t}) \ll 1$, because it takes into account only the modes with very small wavenumbers. The large scales square fluctuations $\left< \dot\sigma^2 \right>$, are given by
\begin{equation}
\left.\left< \dot\sigma^2 \right>  \right|_{y \ll 1} \simeq  \int^{\epsilon k_0(\hat{t})}_0 \, \frac{dk}{k} {\cal P}_{\dot\sigma}(k,\hat{t}),
\end{equation}
where the power-spectrum on cosmological scales: $k\ll k_0(\hat{t})=\beta \hat{t}^{p-1} \left( \frac{9}{4} p^2 -\frac{15}{2} p +2
\right)^{\frac{1}{2}} $, is
\begin{equation}
{\cal P}_{\dot\sigma}(k,\hat{t})  =   \frac{1}{2\pi^2}  \frac{k^{3-2\nu}}{\pi (p-1) (\beta \hat{t})^2}  \,
\left[ \Gamma(\nu_1) \left[2(p-1) \beta\right]^{\nu_1} + (1-3p) \beta \Gamma(\nu) \, \left[2(p-1) \beta\right]^{\nu} \right]^2.
\end{equation}

\subsection{Large-scale solitonic spectrum of $\sigma$ and dark energy}

In order to associate large-scales back-reaction effects with dark energy, we can use the expression (\ref{var}). By requiring $\delta{\cal I}=0$ in (\ref{aac}), we obtain
\begin{eqnarray}
{V}(\hat{t}) &=& \hat{V}_0\, e^{\left[\sigma_0 - \sigma(\hat{t})\right]}, \label{a1} \\
 \left[\frac{\hat{R}}{2\kappa} + \hat{{\cal L}}\right]& = &  \left[\frac{\hat{R}}{2\kappa} + \hat{{\cal L}}\right]_0\, e^{-\left[\sigma_0 - \sigma(\hat{t})\right]}, \label{a2}
\end{eqnarray}
where $\sigma \equiv \left<B\left| \sigma^2\right|B\right>^{1/2}$, $\sigma_0 \equiv \left.\left<B\left| \sigma^2\right|B\right>^{1/2}\right|_{\hat{t}=t_0}$, and
\begin{equation}
\left<B\left| \sigma^2\right|B\right> = \frac{1}{(2\pi)^3} \int^{\epsilon k_0}_0 \, d^3k\, \left|\xi_k\right|^2 \sim \lambda^{n_s-1}, \label{a3}
\end{equation}
where the subscript $0$ denotes the values when inflation starts, $\nu = { 3p -1\over 2 (p-1)}$, and $n_s = 2(\nu-1)\simeq 0.96$ is the spectral index which is measured\cite{pdg} and can be related to the power of the expansion by $p=1+\frac{2}{1-n_s}$, so that $n_s=0.96$ implies that $p=51$\cite{in}. Of course, if the universe is
in a state of accelerated expansion, one must require that $\sigma_0 \geq \sigma(\hat{t})$, for $t\geq t_0$. Therefore, from the expressions (\ref{a1}), (\ref{a2}) and (\ref{a3}), we infer that
the expansion of the universe is produced by some kind of energy, which is coherent at very large scale and produces geometrical fluctuations of the space-time that become smaller at very
large scales, like a manifestation of the expansion.

In order to study the properties of the energy density fluctuations due to the large-scale back-reaction we must calculate the time derivatives of the temporal modes: $\dot{\xi}_{k}(\hat{t})$. They are complex-valued functions, so we can write
\begin{equation}
\dot{\xi}_k(\hat{t}) = \left|\dot{\xi}_k(\hat{t})\right| e^{ik( |\vec{\hat{x}}|_{0} \pm \left| v_{k}(\hat{t}) \right| \hat{t})} ,
\end{equation}
where $\left|\dot{\xi}_k(\hat{t})\right|$ is the norm of $\dot{\xi}_k(\hat{t})$, and $\left| v_{k}(\hat{t}) \right|$ is the norm of the shift of phase-velocity for some modes with
wavenumber norm $k$. If we define $|\vec{\hat{x}}|_{0}=0$ when $\hat{t}=\hat{t}_{0}$ is the time for which inflation begins, we can define the shift phase: $\Omega_k(\hat{t})$, of each $k$-mode, as\footnote{A similar analysis was developed in\cite{*}, but using spherical coordinates to describe pre-inflationary back-reaction effects.}
\begin{equation}
\Omega_k(\hat{t})=\arccos\left(\Re\left[\frac{\dot{\xi}_{k}(\hat{t})}{\left|\dot{\xi}_{k}(\hat{t})\right|}\right]\right)-\arccos\left(\Re\left[\frac{\dot{\xi}_{k}(\hat{t}_{0})}{
\left|\dot{\xi}_{k}(\hat{t}_{0})\right|}\right]\right),
\end{equation}
which describes the degree of coherence of the modes at certain scales, with values close to $k$. If $\Omega_k(\hat{t})$ remains very small with the expansion in a small range of $k$, it will means that the modes remain coherent in this range of the spectrum. We are interested in inspecting what happens on very large scales (i.e. the infrared sector).

\section{Final comments}

\begin{figure}[h]
\noindent
\includegraphics[width=.6\textwidth]{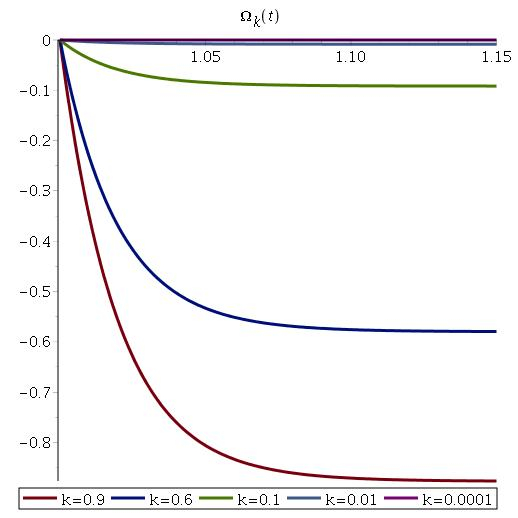}\vskip -0cm\caption{Plot of the shift phase $\Omega_k({\hat{t}})$, for different wavenumbers of the astrophysical and cosmological sector of the spectrum. Notice that the modes with smaller $k$, that describes the cosmological (large-scale) sector of the back-reaction spectrum during inflation, remain almost coherent. However, for smaller scales the coherence between these modes is lost.}\label{f1}
\end{figure}

When inflation begins, the modes of the back-reaction field $\sigma$ are all in phase: $\Omega_k(\hat{t}=\hat{t}_{0})=0$, and therefore all the modes are initially considered as coherent. In the Fig. (\ref{f1}) we have plotted shift phase with different wavenumbers $k$, which correspond to the astrophysical and cosmological sector of the spectrum. Notice that the modes with bigger $k$ (i.e. those of the ultraviolet sector of the back-reaction spectrum), are more shifted than those with smaller $k$ (infrared sector). The important fact is that modes with super-Hubble wavelengths have phase-velocities very similar, and therefore the shift phase of these modes is close to zero: $\left.\Omega_k(\hat{t}=\hat{t}_{0})\right|_{k\rightarrow 0}\rightarrow 0$. This means that those perturbations corresponding to the cosmological (infrared) sector of the spectrum, remain coherent (between them), because there is no tear between these super-Hubble modes during power-law inflation. Of course, it exists a tearing between the group of super-Hubble modes and the astrophysical ones. This means that the decoherence\cite{prd2001} of super-Hubble fluctuations must occur at intermediate scales (i.e. at the border between the infrared and ultraviolet sectors), because the origin of decoherence resides in the interchange of degrees of freedom between the ultraviolet (short-wavelength) and the infrared (super-Hubble scales) sectors. For this reason, the group of very large-scales modes can be considered to have solitonic properties, associated to the extreme of the spectrum of the coarse-grained geometrical field, which describes back-reaction effects at cosmological scales. This should be a possible explanation of the present day observed large scale homogeneity and isotropy of energy density fluctuations, and perhaps more important, for the origin of dark energy, which is responsible for the accelerated expansion of the universe.

\begin{figure}[h]
\noindent
\includegraphics[width=.6\textwidth]{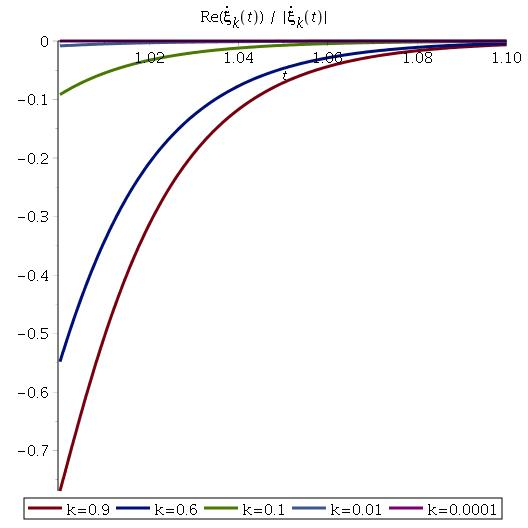}\vskip -0cm\caption{Plot of ${\bf Re}{\left[\dot{\xi}_k\over \left|\dot{\xi}_{k}(\hat{t})\right|\right]}$ for different values of $k$, which becomes negligible with the expansion of the universe.}\label{f2}
\end{figure}

\begin{figure}[h]
\noindent
\includegraphics[width=.6\textwidth]{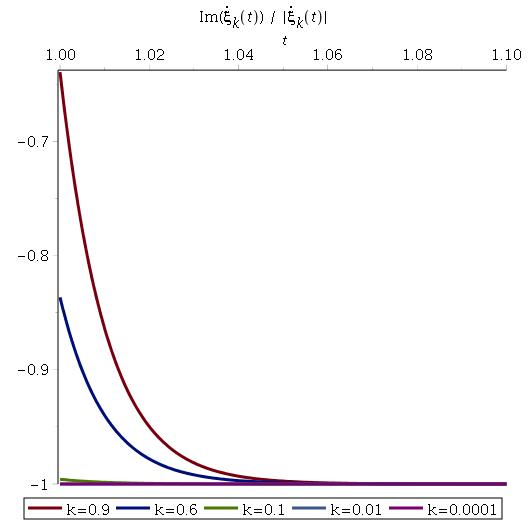}\vskip -0cm\caption{Plot of ${\bf Im}{\left[\dot{\xi}_k\over \left|\dot{\xi}_{k}(\hat{t})\right|\right]}$ for different values of $k$. Notice that ${\dot{\xi}_k\over \left|\dot{\xi}_{k}(\hat{t})\right|}$ becomes totally imaginary as one expects in a quantum-to-classical transition of $\dot{\xi}_k $.}\label{f3}
\end{figure}

This effect is accompanied with a quantum-to-classical transition of the modes that cross from the ultraviolet to the infrared sector. It can be seen in the figures (\ref{f2}) and (\ref{f3}), in which is evident that the time derivative of the modes rotate in the complex plane to finally lie on the imaginary axis. The quantum-to-classical transition of the inflaton field fluctuations is well known\cite{prd2001}. However, this case is really more interesting because is accompanied by a solitonic behavior of the very large scale modes, which remain in phase.


\section*{Acknowledgements}

\noindent The authors acknowledge CONICET, Argentina (PIP 11220150100072CO) and UNMdP (EXA750/16), for financial support.

\newpage
\end{document}